\documentclass[amsmath,prb,twocolumn,superscriptaddress]{revtex4}
\usepackage{color}
\usepackage{amsfonts}
\usepackage{amssymb}
\usepackage{graphicx}
\usepackage{pstricks}
\usepackage{braket}

\begin{document}

\title{Markovian Treatment of non-Markovian Dynamics of Open Fermionic Systems} 

\author{Feng Chen}
\affiliation{Department of Physics, University of California San Diego, La Jolla, CA 92093, USA}
\author{Enrico Arrigoni}
\affiliation{Institute of Theoretical and Computational Physics, Graz University of Technology, 8010 Graz, Austria}
\author{Michael Galperin}
\email{migalperin@ucsd.edu}
\affiliation{Department of Chemistry \& Biochemistry, University of California San Diego, La Jolla, CA 92093, USA}

\begin{abstract}
We show that an open fermionic system coupled to continuous environment with unitary system-environment evolution
can be exactly mapped onto an auxiliary system consisting of the physical fermion system and a set of discrete
fermionic modes subject to non-unitary Lindblad-type system-modes evolution in such a way 
that reduced dynamics of the fermionic system in the two cases are the same. 
Conditions for equivalence of reduced dynamics in the two systems are identified and a proof is presented.
The study is extension of recent work on Bose systems 
[D.~Tamascelli, A.~Smirne, S.~F.~Huelga, and M.~B.~Plenio, Phys. Rev. Lett. \textbf{120}, 030402 (2018)]
to open quantum Fermi systems and to multi-time correlation functions. 
Numerical simulations within generic junction model are presented for  illustration.
\end{abstract}

\maketitle

\section{Introduction}
Open nonequilibirum systems are at the forefront of experimental and theoretical research due to rich and complex physics
they provide access to as well as due to applicational prospects of building nanoscale devices for quantum based technologies
and computations~\cite{jiang_repetitive_2009,khasminskaya_fully_2016,gaita-arino_molecular_2019}. 
Especially intriguing in term of both fundamental science and potential applications are effects of strong correlations. 
A number of impurity solvers capable of treating strongly correlated systems coupled to continuum of 
baths degrees of freedom were developed. Among them are numerical renormalization group in the basis of scattering 
states~\cite{anders_steady-state_2008,schmitt_comparison_2010}, 
flow equations~\cite{wegner_flow-equations_1994,kehrein_flow_2006},
time-dependent density matrix renormalization group~\cite{schollwock_density-matrix_2005,schollwock_density-matrix_2011}, 
multilayer multiconfiguration time-dependent Hartree (ML-MCTDH)~\cite{wang_numerically_2009,wang_multilayer_2018},
and continuous time quantum Monte Carlo~\cite{cohen_taming_2015,antipov_currents_2017,ridley_numerically_2018} 
approaches. These numerically exact techniques are very demanding and so far are mostly applicable to simple models only.

At the same time, accurate numerically inexpensive impurity solvers are in great demand both as standalone techniques
to be applied in simulation of, e.g., nanoscale junctions and as a part of divide-and-conquer schemes such as, e.g, 
dynamical mean-field theory (DMFT)~\cite{anisimov_electronic_2010,aoki_nonequilibrium_2014}. 
In this respect ability to map complicated non-Markovian dynamics of a system onto
much simpler Markov consideration is an important step towards creating new computational techniques
applicable in realistic simulations. In particular, such mapping was used in auxiliary master equation approach (AMEA)~\cite{arrigoni_nonequilibrium_2013,dorda_auxiliary_2014}
introducing numerically inexpensive and pretty accurate solver for the nonequilibrium DMFT.
Another example is recent formulation of the auxiliary dual-fermion method~\cite{chen_auxiliary_2019}.
While the mappings appear to be very useful and accurate, only semi-quantitative arguments for possibility of the mapping
were presented with main supporting evidence being benchmarking vs. numerically exact computational techniques.
In particular, a justification for the mapping was argued in Refs.~\cite{schwarz_lindblad-driven_2016,dorda_optimized_2017,Arrigoni2018} based upon the singular coupling derivation of the Lindblad equation. 
Still, the consideration is not rigorous.

Recently, a rigorous proof of non-Markov to Markov mapping for open Bose quantum systems was presented in the literature~\cite{tamascelli_nonperturbative_2018}.
It was shown that evolution of  reduced density matrix in non-Markov system with unitary system-environment evolution can be
equivalently presented by Markov evolution of an extended system (system plus modes of environment) 
under non-unitary (Lindblad-type) evolution. Here, we extend consideration of Ref.~\cite{tamascelli_nonperturbative_2018} 
to fermionic open quantum systems and to multi-time correlation functions. 
The structure of the paper is the following. After introducing physical and auxiliary models of an open quantum Fermi
system in Section~\ref{model} we discuss non-Markov to Markov mapping procedure
in Section~\ref{map}. Exact mathematical proofs are given in Appendices. Section~\ref{numres} presents numerical illustration of
the mapping for a simple generic model of a junction. We conclude in Section~\ref{conclude}.


\begin{figure}[b]
\centering\includegraphics[width=\linewidth]{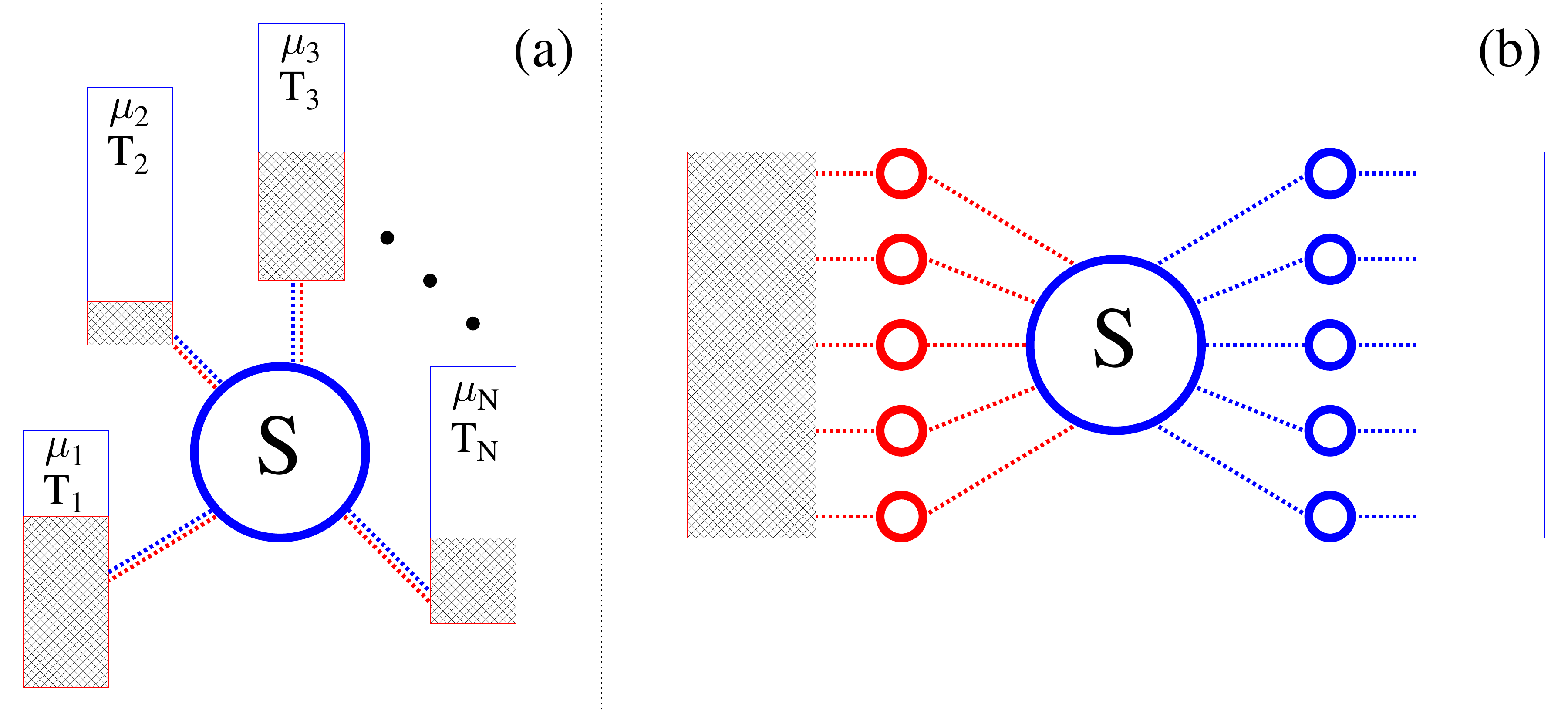}
\caption{\label{fig1}
Sketch of an open fermionic system $S$. Shown are
(a) physical system coupled to $N$ baths and
(b) illustration for an auxiliary system with coupling to full (left) and empty (right) baths.
}
\end{figure}


\section{Models}\label{model}
We consider an open fermionic system $S$ coupled to an arbitrary number $N$ of external baths,
initially each at its own thermodynamic equilibrium, i.e. characterized by its own electrochemical potential and temperature 
(see Fig.~\ref{fig1}a).  The Hamiltonian of the model is
\begin{equation}
\label{Hphys}
\hat H^{phys}(t) = \hat H_S(t) + \sum_{B=1}^{N}\bigg( \hat H_{B} + \hat V_{SB} \bigg)
\end{equation}
Here $\hat H_S(t)$ and $\hat H_B$ ($K\in\{1,\ldots,N\}$) are Hamiltonians of the system and baths.
$\hat V_{SB}$ introduces coupling of the system to bath $B$.
While the Hamiltonian of the system is general and may be time-dependent, we follow the usual paradigm
by assuming bi-linear coupling in constructing fermionic junction models.
\begin{eqnarray}
 \hat H_B &= \sum_{k\in B} \varepsilon_{B\, k}\hat c_{Bk}^\dagger\hat c_{Bk}
 \\
 \hat V_{SB} &= \sum_{k\in B} \sum_{i\in S}\bigg( V_{i,Bk} \hat d_i^\dagger\hat c_{Bk} + H.c. \bigg)
\end{eqnarray}
where $\hat d_i^\dagger$ ($\hat d_i$) and $\hat c_{Bk}^\dagger$ ($\hat c_{Bk}$) create (annihilate)
electron in level $i$ of the system $S$ and level $k$ of bath $B$.
In the model, dynamics of the system-plus-baths evolution is unitary. Below we call this model $phys$ (physical).
We note in passing that extension of the consideration to other types of system-baths couplings is straightforward,
as long as baths are quadratic in the Fermi operators.

The other configuration we'll consider is a model we shall call $aux$ (auxiliary; see Fig.~\ref{fig1}b). Here, the same system $S$
is coupled to a number of auxiliary modes $A$, which in their turn are coupled to two baths. There are two Fermi baths in the
configuration: one ($L$) is completely full ($\mu_L\to+\infty$), the other  ($R$) is completely empty ($\mu_R\to-\infty$).
The Hamiltonian of the system is
\begin{equation}
\hat H^{aux}(t) = \hat H_{S}(t) + \hat V_{SA} + \hat H_A + \sum_{C=L,R}\bigg( \hat H_C + \hat V_{AC} \bigg) 
\end{equation}
where $\hat H_S$ is the same as in (\ref{Hphys}), $\hat H_A$ represents set of modes 
\begin{equation}
 \hat H_A = \sum_{m_1,m_2\in A} H^{A}_{m_1m_2}\hat a_{m_1}^\dagger\hat a_{m_2}
\end{equation}
and $\hat V_{SA}$ their interaction with the system
\begin{equation}
 \hat V_{SA} = \sum_{i\in S}\sum_{m\in A}\bigg( V^{SA}_{i m}\hat d_i^\dagger\hat a_m + H.c. \bigg)
\end{equation}
Here $\hat a_m^\dagger$ ($\hat a_m$) creates (annihilates) electron in the auxiliary mode $m$ in $A$.

$\hat H_C$ represents continuum of states in contact $C$ 
\begin{equation}
 \hat H_C = \sum_{k\in C}\varepsilon_{Ck} \hat c_{Ck}^\dagger\hat c_{Ck}
\end{equation}
with constant density of states
\begin{equation}
 N_C(E) \equiv \sum_{k\in C} \delta(E-\varepsilon_{Ck}) = const
\end{equation}
and $\hat V_{AC}$ couples auxiliary modes $A$ to bath $C$ ($L$ or $R$)
\begin{equation}
\hat V_{AC} = \sum_{k\in C} \sum_{m\in A}\bigg(  t^C_m\hat a^\dagger_m \hat c_{Ck} + H.c. \bigg)
\end{equation}
Dynamics of the whole configuration is unitary.

In the next section we show that the reduced time evolution of $S$ in models $phys$ and $aux$ is the same 
(subject to certain conditions)
and that the reduced dynamics of $S+A$ in model $aux$ satisfies an appropriate Lindblad Markov evolution.
This establishes procedure for Markov non-unitary Lindblad-type treatment of $S+A$ in $aux$
exactly representing unitary non-Markov dynamics of $S$ in $phys$ by tracing out $A$ degrees of freedom. 


\section{Non-Markov to Markov mapping}\label{map}
First, we are going to prove that with an appropriate choice of parameters of $aux$ the dynamics of $S$ 
can be equivalently represented in the original model $phys$ 
and auxiliary model $aux$, under assumption that the dynamics of the whole system is unitary.
Because non-interacting baths are fully characterized by their two-time correlation functions,
equivalence of system-bath(s) hybridizations (i.e. correlation functions of the bath(s) dressed with system-bath(s) interactions)
for the two models indicates equivalence of the reduced system dynamics in the two cases. 
For example, hybridization function is the only information about baths in numerically exact 
simulations of strongly correlated systems~\cite{antipov_currents_2017}. Nonequilibrium character of the system
requires fitting two projections of the hybridization function (also called self-energy in the literature). In particular, these may be 
retarded and Keldysh projections. Let $\Sigma_B^r(E)$ and $\Sigma_B^K(E)$ are matrices
introducing the corresponding hybridization functions
for bath $\alpha$ of the physical problem (Fig.~\ref{fig1}a). 
\begin{eqnarray}
  \big(\Sigma_B^r(E)\big)_{ij} &= \sum_{k\in B} V_{i,Bk}\, g^r_{Bk}(E)\, V_{Bk,j}
  \\
 \big(\Sigma_B^K(E)\big)_{ij}&= \sum_{k\in B} V_{i,Bk}\, g^K_{Bk}(E)\, V_{Bk,j}
\end{eqnarray}
where $g^{r\, (K)}_{Bk}(E)$ are the Fourier transforms of retarded (Keldysh) projections of the free
electron Green's function $g_{Bk}(\tau,\tau')=-i\langle T_c\,\hat c_{Bk}(\tau)\,\hat c_{Bk}^\dagger(\tau')\rangle$.
Then total hybridization functions for the system
\begin{eqnarray}
 \Sigma^r(E) &= \sum_{B=1}^{N} \Sigma_B^r(E)
 \\
 \Sigma^K(E) &= 2\, i\sum_{B=1}^{N} \bigg(1-2f_B(E)\bigg)\mbox{Im}\,\Sigma^r_B(E)
\end{eqnarray}
should be identical with the corresponding hybridization functions, $\tilde\Sigma^r(E)$ and $\tilde\Sigma^K(E)$,
of $S$ in the auxiliary model (Fig.~\ref{fig1}b). The latter have contribution from full ($L$) and empty ($R$) baths,
and from auxiliary modes ($A$)
\begin{eqnarray}
 \tilde\Sigma^r(E) &= \tilde\Sigma^r_L(E) + \tilde\Sigma^r_R(E)
 \\
 \tilde\Sigma^K(E) &= 2\,i\,\mbox{Im}\bigg(\tilde\Sigma^r_R(E)-\tilde\Sigma^r_L(E)\bigg)
\end{eqnarray}
where we assume modes $A$ initially in stationary state.
Requirement of equivalence can be expressed as
\begin{eqnarray}
\mbox{Im}\,\tilde\Sigma^r_L(E) &= \frac{2\, i\,\mbox{Im}\Sigma^r(E)+\Sigma^K(E)}{4\,i}
\\
\mbox{Im}\,\tilde\Sigma^r_R(E) &= \frac{2\, i\,\mbox{Im}\Sigma^r(E)-\Sigma^K(E)}{4\,i}
\end{eqnarray}
Thus, the problem reduces to fitting known functions in the right side of the expression with multiple contributions
from auxiliary modes to the hybridization functions in the left side.
In principle, this fitting can be done in many different ways~\cite{dorda_optimized_2017}. 
For example, possibility of exact fitting of an arbitrary function with
set of Lorentzians was discussed in Ref.~\cite{imamoglu_stochastic_1994}. 
In auxiliary systems such fitting corresponds to 
a construction where each auxiliary mode is coupled to its own bath.
Note that in practical simulations accuracy of the results can be improved either by increasing number of auxiliary modes, 
as is implemented in, e.g, AMEA~\cite{dorda_auxiliary_2015}, or by employing diagrammatic expansion related to the difference
between true and fitted hybridization functions,
as is realized in, e.g., dual fermion approach~\cite{jung_dual-fermion_2012}, or both. 

Now, when equivalence of reduced system ($S$) dynamics in $phys$ and $aux$ is established, we turn to
consideration of evolution of the $aux$ model. We will show that reduced $S+A$ dynamics derived from unitary evolution 
of the $aux$ model can be exactly represented by non-unitary Lindblad-type evolution.

Following Ref.~\cite{tamascelli_nonperturbative_2018} we consider reduced density operator of $S+A$ in $aux$,
$\hat \rho_{SA}$, which is defined by integrating out baths degrees of freedom of the total density operator 
$\hat\rho^{aux}(t)$
\begin{equation}
 \hat\rho_{SA}(t)\equiv \mbox{Tr}_{LR}\bigg[\hat\rho^{aux}(t)\bigg]
\end{equation}
The latter follows unitary evolution with initial condition being $S+A$ decoupled from the baths
\begin{equation}
\label{rhoaux0}
 \hat\rho^{aux}(0) = \hat\rho_L\otimes\hat\rho_{SA}(0)\otimes\hat\rho_R
\end{equation}
where $\hat\rho_L=\vert full\rangle\langle full\vert$ and $\hat\rho_R=\vert empty\rangle\langle empty\vert$.

In Appendix~\ref{appA} we prove that $\hat\rho_{SA}(t)$ satisfies Markov Lindblad-type equation of motion
\begin{align}
\label{rho_EOM}
&\frac{d}{dt} \hat\rho_{SA}(t) = -i\bigg[\hat H_{SA}(t),\hat \rho_{SA}(t)\bigg] + \sum_{m_1,m_2\in A}\bigg[
\nonumber \\ & \quad
\Gamma^R_{m_1m_2}\bigg(\hat a_{m_2}\hat\rho_{SA}(t)\hat a_{m_1}^\dagger-\frac{1}{2}\left\{\hat\rho_{SA}(t),\hat a_{m_1}^\dagger\hat a_{m_2}\right\}\bigg)
\\ &
+\Gamma^L_{m_1m_2}\bigg(\hat a_{m_1}^\dagger\hat\rho_{SA}(t)\hat a_{m_2}-\frac{1}{2}\left\{\hat\rho_{SA}(t),\hat a_{m_2}\hat a_{m_1}^\dagger\right\}\bigg)
\bigg]
\nonumber \\ 
&\equiv \mathcal{L}_{SA}(t) \vert\rho_{SA}(t)\rangle\rangle
\nonumber
\end{align}
where 
\begin{equation}
\hat H_{SA}(t)\equiv \hat H_S(t)+\hat V_{SA}+\hat H_A,
\end{equation}
$\mathcal{L}_{SA}$ is the Liouvillian superoperator defined on the $S+A$ subspace of the $aux$ model and 
\begin{equation}
\label{Gammadef}
 \Gamma^C_{m_1m_2}\equiv 2\pi t^C_{m_1} (t_{m_2}^{C})^{*} N_C \qquad (C=L,R)
\end{equation}
is the dissipation matrix due to coupling to contact $C$.


\begin{figure}[htbp]
\centering\includegraphics[width=0.8\linewidth]{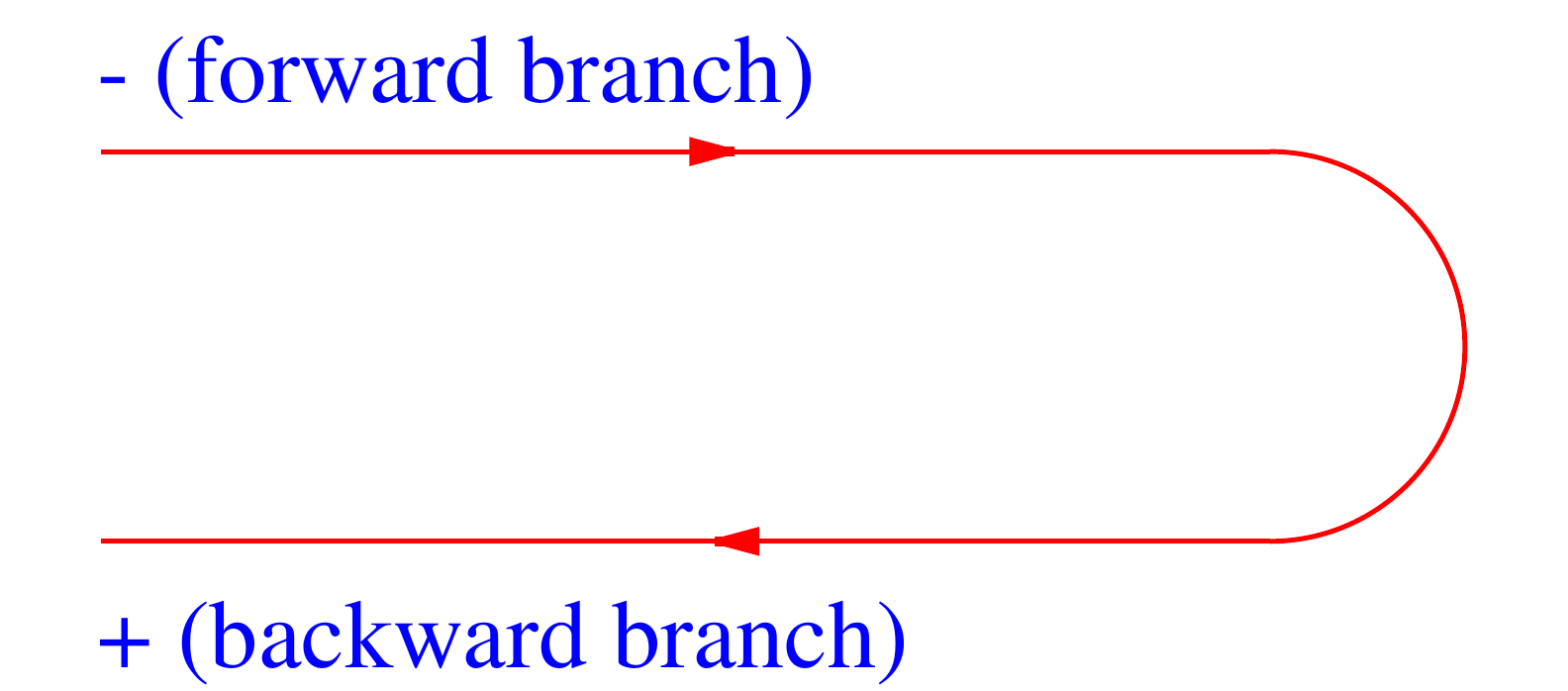}
\caption{\label{fig2}
The Keldysh contour. 
}
\end{figure}


Next we turn to multi-time correlation functions of operators in the $S+A$ subspace of the $aux$ model.
Following Ref.~\cite{tamascelli_nonperturbative_2018} we start consideration from two-time correlation function on real time axis.
For arbitrary operators $\hat O_1$ and $\hat O_2$ in $S+A$ we define two-time ($t_1\geq t_2\geq 0$)
correlation function as 
\begin{align}
 \label{corr2rt}
 & \langle\hat O_1(t_1)\hat O_2(t_2)\rangle \equiv 
 \\ &
\mbox{Tr}\bigg[\hat O_1\,\hat U^{aux}(t_1,t_2)\,\hat O_2\,\hat U^{aux}(t_2,0)\,\hat \rho^{aux}(0)\,\hat U^{aux\,\dagger}(t_1,0)\bigg]
\nonumber
\end{align}
Here $\hat U^{aux}$ is the evolution operator in the $aux$ system
\begin{equation}
 \label{Uaux}
 \hat U^{aux}(t,t') \equiv T \exp\bigg[-i\int_{t'}^{t}ds\, \hat H^{aux}(s)\bigg]
\end{equation}
and $T$ is the time-ordering operator.
In \ref{appB} we show that (\ref{corr2rt}) can be equivalently obtained from reduced Linblad-type evolution in 
the $S+A$ subspace
\begin{align}
 \label{corr2rt_EOM}
&\langle\hat O_1(t_1)\hat O_2(t_2)\rangle = 
\\ &\quad
\langle\langle I\vert \mathcal{O}^{-}_1\,\mathcal{U}_{SA}(t_1,t_2)\,
\mathcal{O}^{-}_{2}\,\mathcal{U}_{SA}(t_2,0)\,\vert\rho_{SA}(0)\rangle\rangle
\nonumber
\end{align}
Here $\langle\langle I\vert$ is Liouville space bra representation of the Hilbert space identity operator, 
$\mathcal{O}_i$ is the Liouville space superoperator corresponding to the Hilbert space operator $\hat O_i$
(see Fig.~\ref{fig2})
\begin{equation}
\label{superop}
 \mathcal{O}_i \vert \rho\rangle\rangle = \left\{
 \begin{array}{cc}
  \mathcal{O}_i^{-} \vert \rho\rangle\rangle \equiv \hat O_i\,\hat\rho & \mbox{forward branch} 
  \\
  \mathcal{O}_i^{+} \vert \rho\rangle\rangle \equiv \hat\rho\,\hat O_i & \mbox{backward branch} 
 \end{array}
 \right.
\end{equation}
and $\mathcal{U}_{SA}$ is the Liouville space evolution superoperator
\begin{equation}
\label{ULiouv}
 \mathcal{U}_{SA}(t,t') \equiv T \exp\bigg[\int_{t'}^t ds\,\mathcal{L}_{SA}(s)\bigg]
\end{equation}

Finally, we extend consideration to multi-time correlation functions of arbitrary operators $\hat O_i$ ($i\in\{1,\ldots,N\}$)
defined on the Keldysh contour (see Fig.~\ref{fig2}) as
\begin{align}
\label{corrN}
& \langle T_c\, \hat O_1(\tau_1)\,\hat O_2(\tau_2)\ldots\hat O_N(\tau_N)\rangle
 \equiv
 \\ &\qquad
  \mbox{Tr}\bigg[T_c\,\hat O_1\,\hat O_2\ldots\hat O_N\,\hat U_c\,\hat\rho^{aux}(0)\bigg]
  \nonumber
\end{align}
where $\tau_i$ are the contour variables, $T_c$ is the contour ordering operator, and
\begin{equation}
 \hat U_c = T_c \exp\bigg[-i\int_c d\tau\,\hat H^{aux}(\tau)\bigg]
\end{equation}
is the contour evolution operator. 
Note subscripts of operators $O_i$ in the right side of (\ref{corrN}) indicate both type of the operator and its position on the contour.
In \ref{appC} we prove that multi-time correlation functions (\ref{corrN}) can be evaluated solely from
Markov Lindblad-type evolution in $S+A$ subspace of the $aux$ model
\begin{align}
\label{corrN_EOM}
 &\langle T_c\, \hat O_1(\tau_1)\,\hat O_2(\tau_2)\ldots\hat O_N(\tau_N)\rangle
 =
 \\ &\quad (-1)^P
 \langle\langle I\vert \mathcal{O}_{\theta_1}\,\mathcal{U}_{SA}(t_{\theta_1},t_{\theta_2})\, \mathcal{O}_{\theta_2}\,
 \mathcal{U}_{SA}(t_{\theta_2},t_{\theta_3})\ldots 
\nonumber \\ &\qquad\qquad\qquad\qquad\quad\ldots
 \mathcal{O}_{\theta_N}\mathcal{U}_{SA}(t_{\theta_N},0)
 \vert \rho_{SA}(0)\rangle\rangle
 \nonumber
\end{align}
 Here $P$ is number of Fermi interchanges in the permutation of operators $\hat O_i$ by $T_c$, 
 $\theta_i$ are indices of operators $\hat O_i$ rearranged is such a way that 
 $t_{\theta_1}>t_{\theta_2}>\ldots >t_{\theta_N}$
 ($t_{\theta_i}$ is real time corresponding to contour variable $\tau_{\theta_i}$),
 $\mathcal{O}_{\theta_i}$ are the superoperators defined in (\ref{superop}),
 and $\mathcal{U}_{SA}$ is the Liouville space evolution superoperator defined in (\ref{ULiouv}).
 
 Equivalence of $S$ dynamics derived from unitary evolution of models $phys$ and $aux$
 together with (\ref{rho_EOM}) and (\ref{corrN_EOM}) completes proof of possibility of Markov treatment for 
 non-Markovian dynamics in open quantum Fermi systems.

\section{Numerical illustration}\label{numres}


\begin{figure}[htbp]
\centering\includegraphics[width=\linewidth]{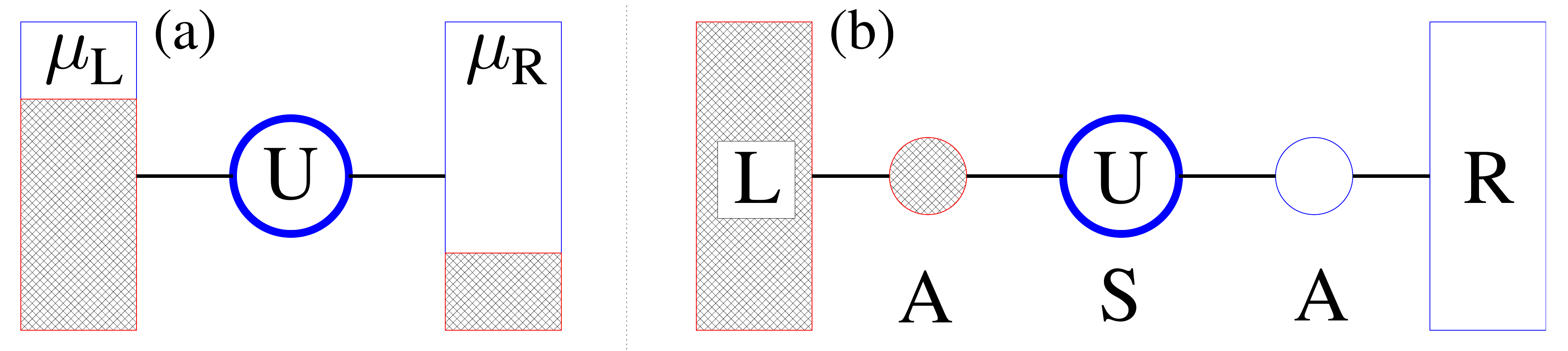}
\caption{\label{fig3}
Original Anderson impurity (a) and corresponding auxiliary (b) models. 
}
\end{figure}



\begin{figure}[htbp]
\centering\includegraphics[width=0.6\linewidth]{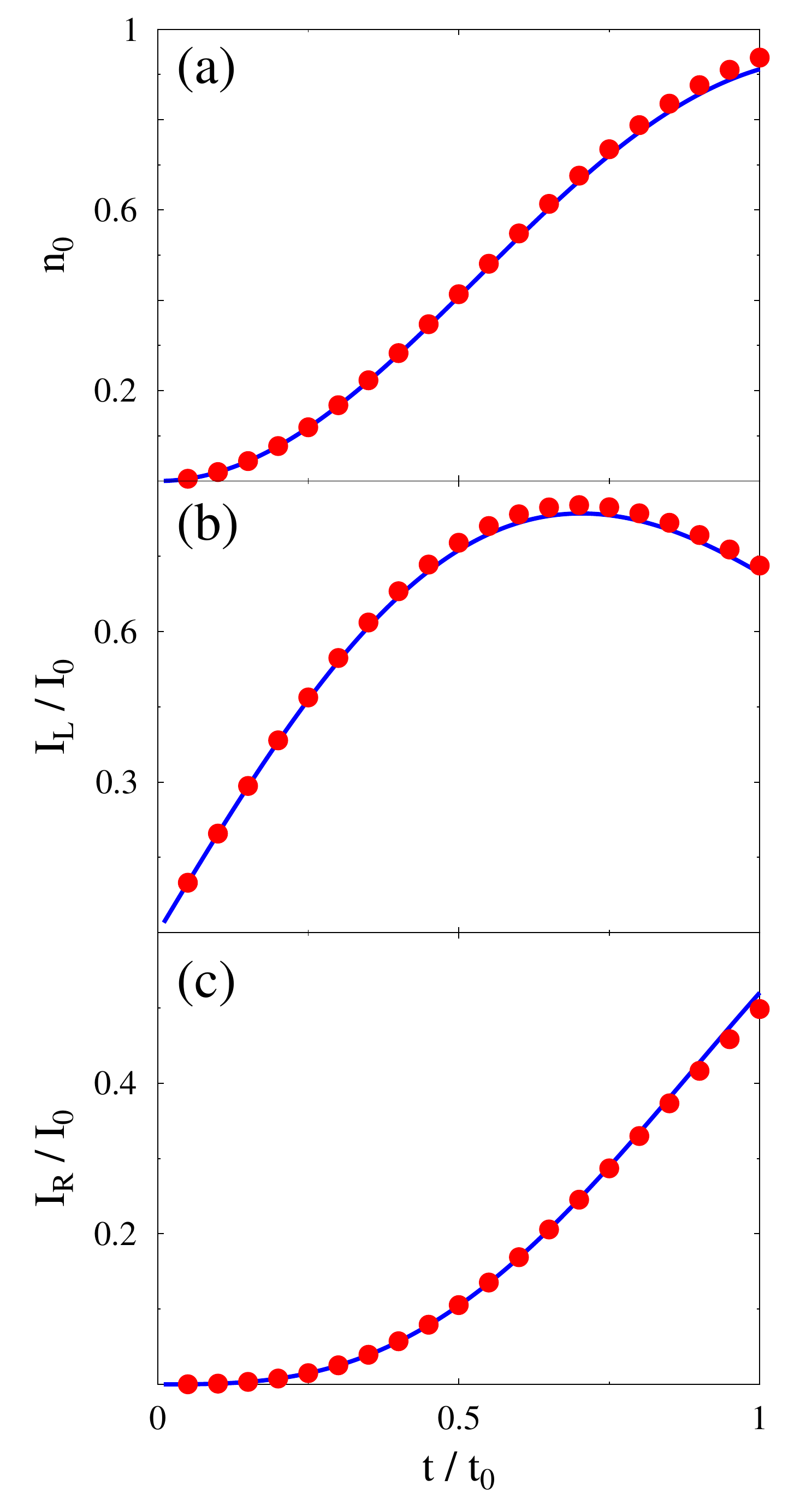}
\caption{\label{fig4}
Unitary (filled circles, red) and Lindblad-type (solid line, blue) evolution in auxiliary model of Fig.~\ref{fig3}b
after connecting initially empty central site to filled $L$ and empty $R$ baths.
Shown are population of the level (a) and left (b) and right (c) currents. 
See text for parameters.
}
\end{figure}

Here we present numerical simulation illustrating equivalence of original unitary and 
Lindblad-type Markov treatment for the open quantum Fermi system.
We consider Anderson model (Fig.~\ref{fig3}a)
\begin{align}
 \hat H &=\sum_{\sigma\in\{\uparrow,\downarrow\}}\varepsilon_0\hat d_\sigma^\dagger\hat d_\sigma + U \hat n_\uparrow\hat n_\downarrow
 \\ &
 +\sum_{k\in L,R}\sum_{\sigma\in\{\uparrow,\downarrow\}}\bigg(\varepsilon_k\hat c_{k\sigma}^\dagger\hat c_{k\sigma}
 + V_k \hat d^\dagger_{\sigma}\hat c_{k\sigma} + V_k^{*} \hat c_{k\sigma}^\dagger \hat d_\sigma\bigg)
 \nonumber
\end{align}
where $\hat n_\sigma=\hat d_\sigma^\dagger\hat d_\sigma$.
We calculate the system evolution after connecting initially empty site to baths at time $t=0$.
Parameters of the simulations are (numbers are in arbitrary units of energy $E_0$):
$\varepsilon_0=0$ and $U=1$.  We assume
\begin{equation}
\Gamma_K(E)=\gamma_K\frac{t_K^2}{(E-\varepsilon_0)^2+(\gamma_K/2)^2}
\end{equation}
where $\Gamma_K(E)\equiv 2\pi\sum_{k\in K}\vert V_k\vert^2\delta(E-\varepsilon_k)$
is the electron escape rate into contact $K$, $\gamma_L=\gamma_R=0.2$, and $t_L=t_R=1$.

For simplicity, we consider high bias, so that auxiliary model with only two sites (Fig.~\ref{fig3}b) 
is sufficient to reproduce dynamics in the physical system. 
In the auxiliary model we compare unitary evolution calculated within numerically exact
td-DMRG~\cite{schollwock_density-matrix_2005,schollwock_density-matrix_2011,bauer_alps_2011,dolfi_matrix_2014} with Lindblad QME results.  Time is shown in units of $t_0=\hbar/E_0$, currents use $I_0=E_0/\hbar$,
and $\hbar$ is assumed to be $1$.
Figure~\ref{fig4} shows level population, $n_0=\langle\hat n_\sigma\rangle$, as well as left, $I_L$, and right, $I_R$, currents
in the system after quench. 
Close correspondence between the two numerical results is an illustration
for exact analytical derivations presented in Section~\ref{map}.

\section{Conclusions}\label{conclude}
We consider open quantum Fermi system $S$ coupled to a number of external Fermi baths each at its own equilibrium
(each bath has its own electrochemical potential $\mu_i$ and temperature $T_i$). 
Evolution of the model (system plus baths) is unitary.
We show that reduced dynamics of the system $S$ in the original unitary non-Markov model can be 
exactly reproduced by Markov non-unitary Lindblad-type evolution of an auxiliary system,
which consists of the system $S$ coupled to a number of auxiliary modes $A$ which in turn are coupled
to two Fermi baths $L$ and $R$: one full ($\mu_L\to+\infty$) and one empty ($\mu_R\to-\infty$).
The proof is performed in two steps: first we show that reduced $S$ dynamics in the physical model  is
equivalent to reduced dynamics of $S$ in the auxiliary model, when $A$ degrees of freedom and the two baths are
traced out; second, we show that reduced dynamics of $S$+$A$ in the auxiliary model with unitary evolution of the model
can be exactly reproduced by the Lindblad-type Markov evolution of $S$+$A$.
The correspondence is shown to hold for reduced density matrix and for multi-time correlation functions defined 
on the Keldysh contour. Our study is extension of recent work about Bose systems~\cite{tamascelli_nonperturbative_2018}
to open Fermi systems and beyond only reduced density matrix consideration.
Establishing possibility of exact mapping of reduced unitary non-Markov dynamics to much simpler
non-unitary Markov Lindbald-type treatment sets firm basis for auxiliary quantum master equations (QME) methods
employed in, e.g, AMEA~\cite{arrigoni_nonequilibrium_2013} or aux-DF~\cite{chen_auxiliary_2019} approaches.
We note that in practical implementations improving quality of mapping can be based on increasing number of
$A$ modes, as is done in advanced AMEA implementations~\cite{dorda_auxiliary_2015}, or by utilization of expansion in 
discrepancy between physical and auxiliary hybridization functions, as is done in the dual fermion 
formulation~\cite{jung_dual-fermion_2012},
or both. Scaling performance of the two approaches to mapping quality enhancement is a goal for future research.

\begin{acknowledgments}
We thank Max Sorantin for useful discussions.
This material is based upon work supported by the National Science Foundation under grant CHE-1565939.
\end{acknowledgments}
\appendix

\section{Derivation of Eq.~(\ref{rho_EOM})}\label{appA}
Here we prove that reduced density matrix of $S+A$ in the $aux$ model satisfies Markov Lindblad-type equation-of-motion (EOM), 
Eq.~(\ref{rho_EOM}).

We start by considering unitary evolution of the $aux$ model. 
Heisenberg EOM for bath annihilation operator $\hat{c}_{Ck}$ is
\begin{align}
\frac{d}{dt}\hat{c}_{Ck}(t)&=i[\hat H^{aux}(t),\hat{c}_{Ck}(t)]
\\ &
 = -i\varepsilon_{Ck}\hat c_{Ck}(t) - i \sum_{m\in A} (t_{m}^{C})^{*}\, \hat a_{m}(t)
 \nonumber
\end{align}
Its formal solution is
\begin{equation}
\hat{c}_{Ck}(t)=e^{-i\varepsilon_{Ck}t}\hat{c}_{Ck}(0)
-i\sum_{m\in A} (t^C_{m})^{*}\int_0^t ds\, e^{i\varepsilon_{Ck}(t-s)}\hat{a}_m(s)
\end{equation}
Thus, Heisenberg EOM for an arbitrary operator $\hat{O}$ on $S+A$ can be written as
\begin{widetext}
\begin{align}
\label{EOM}
&\frac{d}{dt}\hat{O}(t) =i[\hat H_{SA}(t),\hat{O}(t)] -i\sum_{m\in A}  \bigg\{
\\ & \quad
 \sum_{k\in R}\bigg[
t^R_{m} [\hat{O}(t),\hat{a}^\dagger_m]_\zeta
\Big(e^{-i\varepsilon_{Rk}t}\hat{c}_{Rk}(0)-i\sum_{m'\in A} (t^R_{m'})^{*}\int_0^t ds\, e^{-i\varepsilon_{Rk}(t-s)}\hat{a}_{m'}(s)\Big)
\nonumber\\ & \quad
 +\zeta (t^R_{m})^{*}
\Big(e^{i\varepsilon_{Rk}t}\hat{c}^\dagger_{Rk}(0)
      +i\sum_{m'\in A} t^R_{m'}\int_0^t ds\, e^{i\varepsilon_{Rk}(t-s)}\hat{a}^\dagger_{m'}(s)\Big)[\hat{O}(t),\hat{a}_m(t)]_\zeta
      \bigg]
\nonumber\\ &-
 \sum_{k\in L}\bigg[
 (t^L_{m})^*[\hat{O}(t),\hat{a}_m(t)]_\zeta\Big(e^{i\varepsilon_{Lk}t}\hat{c}^\dagger_{Lk}(0)
  +i\sum_{m'\in A} t^L_{m'}\int_0^t ds\, e^{i\varepsilon_{Lk}(t-s)}\hat{a}^\dagger_{m'}(s)\Big)
\nonumber\\ &
 +\zeta t^L_{m} \Big(e^{-i\varepsilon_{Lk}t}\hat{c}_{Lk}(0)
   -i\sum_{m'\in A} (t^L_{m'})^{*}\int_0^t ds\, e^{-i\varepsilon_{Lk}(t-s)}\hat{a}_{m'}(s)\Big)[\hat{O}(t),\hat{a}^\dagger_m(t)]_\zeta
   \bigg] \bigg\}
\nonumber
\end{align}
\end{widetext}
where $\zeta=+/-1$ if $\hat{O}$ contains even/odd number of fermion operators, and $[,]_\zeta$ is (anti)commutator for $\zeta=(-)1$.

For future reference we introduce 
\begin{equation}
\label{cindef}
\hat{c}_C^{(in)}(t)\equiv\sum_{k \in C} e^{-i\varepsilon_{C k}t}\hat{c}_{C k}(0)
\end{equation}
which satisfies usual commutation relations
\begin{align}
 \label{eqin}
	 &\big\{\hat{c}_{C_1}^{(in)}(t),\hat{c}_{C_2}^{(in)\,\dagger}(s)\big\} =\delta_{C_1,C_2}\,\delta(t-s)
	\\
	& \big\{\hat{c}_{C_1}^{(in)}(t),\hat{c}_{C_2}^{(in)}(s)\big\} 
	=\big\{\hat{c}_{C_1}^{(in)\,\dagger}(t),\hat{c}_{C_2}^{(in)\,\dagger}(s)\big\}=0
\end{align}
Note that because contact density of states $N_C$ is constant
\begin{equation} 
 \label{rho}
 \sum_{k\in C} e^{-i\varepsilon_{Ck}t}\equiv \int d\varepsilon\, N_{C}(\varepsilon)e^{-i\varepsilon t}=2\pi N_C\delta(t)
\end{equation}
is satisfied. Note also that 
\begin{equation}
\label{delta}
\int_0^t ds\,\delta(t-s)f(s)=\frac{1}{2}f(t)
\end{equation}
holds for arbitrary function $f(t)$.

Using (\ref{cindef}), (\ref{rho}), and (\ref{delta}) in (\ref{EOM}) leads to
\begin{widetext}
\begin{align}
\label{EOM_in}
\frac{d}{dt}\hat{O}(t)& = i[\hat H_{SA}(t),\hat{O}(t)]
\nonumber\\ & \quad
-i\sum_{m\in A} \bigg\{t_{Rj}[\hat{O}(t),\hat{a}^\dagger_m(t)]_\zeta\,\hat{c}_R^{(in)}(t)
 +\zeta (t^R_{m})^{*}\hat{c}_R^{(in)\,\dagger}(t)[\hat{O}(t),\hat{a}_m(t)]_\zeta
\nonumber \\ & \qquad\qquad
 -(t^L_{m})^{*}[\hat{O}(t),\hat{a}_m(t)]_\zeta\hat{c}_L^{(in)\,\dagger}(t)
 -\zeta t^L_{m}\hat{c}_L^{(in)}(t)[\hat{O}(t),\hat{a}^\dagger_m(t)]_\zeta
 \bigg\}
 \\ & \quad
-\frac{1}{2}\sum_{m_1,m_2\in A} \bigg\{\Gamma^{R}_{m_1m_2}[\hat{O}(t),\hat{a}^\dagger_{m_1}(t)]_\zeta\,\hat{a}_{m_2}(t)
 -\zeta\Gamma^{R}_{m_2m_1}\hat{a}_{m_2}(t)^\dagger[\hat{O}(t),\hat{a}_{m_1}(t)]_\zeta
 \nonumber \\ & \qquad\qquad
   -\zeta\Gamma^{L}_{m_1m_2}\hat{a}_{m_2}(t)[\hat{O}(t),\hat{a}^\dagger_{m_1}(t)]_\zeta
   +\Gamma^{L}_{m_2m_1}[\hat{O}(t),\hat{a}_{m_1}(t)]_\zeta\,\hat{a}_{m_2}^\dagger(t)
   \bigg\}
   \nonumber
\end{align}
where we employed definition of the dissipation matrix, Eq.~(\ref{Gammadef}).

Next we are going to write EOM for expectation value of $\hat O$
\begin{equation}
\braket{\hat{O}(t)}\equiv \mbox{Tr}\big[\hat{O}(t)\,\hat{\rho}^{aux}(0)\big]
\end{equation}
by averaging (\ref{EOM_in}) with initial density operator of the $aux$ model, Eq.~(\ref{rhoaux0}).
Because initially $S+A$ is from the baths and because bath $L$ is full and $R$ is empty (see Fig.~\ref{fig1}b)
\begin{equation}
 \label{kill}
  \hat{c}_L^{(in)\,\dagger}(t)\,\hat\rho_L=\hat\rho_L\,\hat{c}_L^{(in)}(t)=
 \hat{c}_R^{(in)}(t)\,\hat\rho_R=\hat\rho_R\,\hat{c}_R^{(in)\,\dagger}(t)=0
\end{equation}
holds. Thus, second and third lines in (\ref{EOM_in}) do not contribute,
and EOM for the expectation value of $\hat O(t)$  is
\begin{align}
\label{EOM_noin}
&\bigg\langle\frac{d}{dt}\hat{O}(t)\bigg\rangle =\mbox{Tr}\bigg[\hat\rho^{aux}(0)\, i[\hat H_{SA}(t),\hat{O}(t)]\bigg]
-\frac{1}{2}\sum_{m_1,m_2\in A}\mbox{Tr}\bigg[\hat\rho^{aux}(0)
\Big\{
\nonumber \\ & \qquad
\Gamma^R_{m_1m_2}[\hat{O}(t),\hat{a}^\dagger_{m_1}(t)]_\zeta\,\hat{a}_{m_2}(t)
 -\zeta\Gamma^R_{m_1m_2}\hat{a}_{m_1}^\dagger(t)[\hat{O}(t),\hat{a}_{m_2}(t)]_\zeta
 \\ & \quad
 +\Gamma^L_{m_1m_2}[\hat{O}(t),\hat{a}_{m_2}(t)]_\zeta\,\hat{a}_{m_1}^\dagger(t)
 -\zeta\Gamma^L_{m_1m_2}\hat{a}_{m_2}(t)[\hat{O}(t),\hat{a}_{m_1}^\dagger(t)]_\zeta
 \Big\}
 \bigg]
 \nonumber
\end{align}
Because $\hat{O}$ is arbitrary in $S+A$, after transforming to Schr{\" o}dinger picture
(\ref{EOM_noin}) can be rewritten as EOM for $\hat \rho^{aux}(t)$
\begin{align}
\label{EOM_rhoaux}
\frac{d}{dt}\hat\rho^{aux}(t) &=-i[\hat H_{SA}(t),\hat\rho^{aux}(t)]
\nonumber\\
&+\sum_{m_1,m_2\in A}\bigg[
\Gamma^R_{m_1m_2}\bigg(\hat a_{m_2}\hat\rho^{aux}(t)\hat a_{m_1}^\dagger
-\frac{1}{2}\left\{\hat\rho^{aux}(t),\hat a_{m_1}^\dagger\hat a_{m_2}\right\}\bigg)
\\ &\qquad\quad\ \ \ 
+\Gamma^L_{m_1m_2}\bigg(\hat a_{m_1}^\dagger\hat\rho^{aux}(t)\hat a_{m_2}
-\frac{1}{2}\left\{\hat\rho^{aux}(t),\hat a_{m_2}\hat a_{m_1}^\dagger\right\}\bigg)
\bigg]
\nonumber
\end{align}
Finally, because only operators in $S+A$ subspace appear in the right side of (\ref{EOM_rhoaux}),
tracing out baths degrees of freedom leads to Eq.~(\ref{rho_EOM}).
\end{widetext}


\section{Derivation of Eq.~(\ref{corr2rt_EOM})}\label{appB}
Here we prove that two-time correlation function of two arbitrary operators in $S+A$, 
$\langle\hat O_1(t_1)\,\hat O_2(t_2)\rangle$ ($t_1\geq t_2\geq 0$), Eq.~(\ref{corr2rt}), 
can be equivalently obtained from reduced Lindblad-type evolution in the $S+A$ subspace of the $aux$ model.

Let introduce  $t\equiv t_1-t_2 \geq 0$,
then $\hat O_1(t_1)\hat O_2(t_2) = \hat O_1(t+t_2)\hat O_2(t_2)$ and
using Eq.~(\ref{EOM_in}) we get
\begin{widetext}
\begin{align}
\label{Green_in}
&\frac{d}{dt}\hat{O}_1(t+t_2)\hat{O}_2(t_2)  = \bigg\{i[\hat H_{SA}(t+t_2),\hat{O}_1(t+t_2)]
\nonumber \\ &
-i\sum_{m\in A}\Big(
 t^R_{m}[\hat{O}_1(t+t_2),\hat{a}^\dagger_m(t+t_2)]_{\zeta_1}\hat{c}_R^{(in)}(t+t_2)
 +\zeta_1 (t^R_{m})^{*}\hat{c}_R^{(in)\,\dagger}(t+t_2)[\hat{O}_1(t+t_2),\hat{a}_m(t+t_2)]_{\zeta_1}
\nonumber \\& \qquad\quad
 -(t^L_{m})^{*}[\hat{O}_1(t+t_2),\hat{a}_m(t+t_2)]_{\zeta_1}\hat{c}_L^{(in)\,\dagger}(t+t_2)
 -\zeta_1 t^L_{m}\hat{c}_L^{(in)}(t+t_2)[\hat{O}_1(t+t_2),\hat{a}^\dagger_m(t+t_2)]_{\zeta_1}
 \Big)
\\ &
 -\frac{1}{2}\sum_{m_1,m_2\in A}\Big(
 \Gamma^R_{m_1m_2}[\hat{O}_1(t+t_2),\hat{a}^\dagger_{m_1}(t+t_2)]_{\zeta_1}\hat{a}_{m_2}(t+t_2)
 -\zeta_1\Gamma^R_{m_1m_2}\hat{a}_{m_1}^\dagger(t+t_2)[\hat{O}_1(t+t_2),\hat{a}_{m_2}(t+t_2)]_{\zeta_1}
\nonumber \\& \qquad\ \ 
 +\Gamma^L_{m_1m_2}[\hat{O}_1(t+t_2),\hat{a}_{m_2}(t+t_2)]_{\zeta_1}\hat{a}_{m_1}^\dagger(t+t_2)
 -\zeta_1\Gamma^L_{m_1m_2}\hat{a}_{m_2}(t+t_2)[\hat{O}_1(t+t_2),\hat{a}_{m_1}^\dagger(t+t_2)]_{\zeta_1}
 \Big)\bigg\}\hat{O}_2(t_2)
 	\nonumber
\end{align}
\end{widetext}
Note that for $t>0$
\begin{equation}
\label{crucial}
	[\hat{c}_C^{(in)\,\dagger}(t+t_2),\hat{O}_2(t_2)]_{\zeta_2}
	=[\hat{c}_C^{(in)}(t+t_2),\hat{O}_2(t_2)]_{\zeta_2}=0
\end{equation}
Indeed, because from Eq.~(\ref{EOM_in}) it is clear that $\hat{O}_2(t_2)$ depends on $\hat{O}_2(s)$ 
and $\hat{c}_C^{(in)\,(\dagger)}(s)$ only at earlier times ($s<t_2$) and because
Eq.~(\ref{eqin}) shows that  $\hat{c}_C^{(in)\,(\dagger)}(s)$ taken at different times anti-commute 
with each other, Eq.~(\ref{crucial}) holds. 

Thus, while taking the expectation value of (\ref{Green_in}) with respect to $\hat\rho^{aux}(0)$, Eq.~(\ref{rhoaux0}), 
$\hat{c}_L^{(in)\,\dagger}(t+t_2)$ and $\hat{c}_R^{(in)}(t+t_2)$ can be moved over $\hat{O}_2(t_2)$
for any $t>0$. 
So as in \ref{appA}, terms with $\hat{c}_C^{(in)\,(\dagger)}(t) $ in (\ref{Green_in}) 
again don't contribute (see  Eq.~(\ref{kill})), and we get for $t>0$
\begin{widetext}
\begin{align}
\label{EOM_green}
& \frac{d}{dt}\bigg\langle\hat{O}_1(t+t_2)\hat{O}_2(t_2)\bigg\rangle=
\mbox{Tr}\bigg[\bigg\{
i[\hat H_{SA}(t+t_2),\hat{O}_1(t+t_2)] -\frac{1}{2}\sum_{m_1,m_2\in A}\Big(
\nonumber \\ &\quad
 \Gamma^R_{m_1m_2}[\hat{O}_1(t+t_2),\hat{a}^\dagger_{m_1}(t+t_2)]_{\zeta_1}\hat{a}_{m_2}(t+t_2)
 -\zeta_1\Gamma^R_{m_1m_2}\hat{a}_{m_1}^\dagger(t+t_2)[\hat{O}_1(t+t_2),\hat{a}_{m_2}(t+t_2)]_{\zeta_1}
 \\ &
 +\Gamma^L_{m_1m_2}[\hat{O}_1(t+t_2),\hat{a}_{m_2}(t+t_2)]_{\zeta_1}\hat{a}_{m_1}^\dagger(t+t_2)
 -\zeta_1\Gamma^L_{m_1m_2}\hat{a}_{m_2}(t+t_2)[\hat{O}_1(t+t_2),\hat{a}_{m_1}^\dagger(t+t_2)]_{\zeta_1}
 \Big)\bigg\}
 \times \hat{O}_2(t_2)\hat\rho^{aux}(0)
\bigg]
\nonumber
\end{align}
Rearranging evolution operators, Eq.~(\ref{Uaux}), and separating traces over $S+A$ and $L+R$ yields
\begin{align}
\label{rearrange}
& \bigg\langle\hat O_1(t+t_2)\,\hat O_2(t_2)\bigg\rangle = 
 \mbox{Tr}_{SA}\bigg\{ \hat O_1\,\mbox{Tr}_{LR}\bigg[\hat U^{aux}(t+t_2,0)\hat O_2(t_2)\,\hat\rho^{aux}(0)\hat U^{aux\,\dagger}(t+t_2)\bigg]\bigg\}
\\
\label{rearrange_EOM}
&\frac{d}{dt}\bigg\langle\hat{O}_1(t+t_2)\hat{O}_2(t_2)\bigg\rangle  =
\mbox{Tr}_{SA}\bigg\{ \hat O_1\,\frac{d}{dt}\mbox{Tr}_{LR}\bigg[\hat U^{aux}(t+t_2,0)\hat O_2(t_2)\,\hat\rho^{aux}(0)\hat U^{aux\,\dagger}(t+t_2)\bigg]\bigg\}
\end{align}
\end{widetext}
So that (\ref{EOM_green}) can be rewritten as
\begin{equation}
\begin{array}{ll}
\label{EOM_green_rewritten}
\mbox{Tr}_{SA}\bigg\{ \hat O_1\,\frac{d}{dt}\mbox{Tr}_{LR}[\ldots] \bigg\} &=
\mbox{Tr}_{SA}\bigg\{ \bigg(\mathcal{L}_{SA}^{\dagger}(t)\,\hat O_1\bigg) \mbox{Tr}_{LR}[\ldots] \bigg\}
\\ & 
\equiv \mbox{Tr}_{SA}\bigg\{ \hat O_1 \,\mathcal{L}_{SA}(t)\mbox{Tr}_{LR}[\ldots] \bigg\}
\end{array}
\end{equation}
where $\mathcal{L}_{SA}^\dagger(t)$ is adjoint~\cite{breuer_theory_2003} 
of the Liouvillian $\mathcal{L}_{SA}(t)$ defined in (\ref{rho_EOM}),
and where $\mbox{Tr}_{LR}[\ldots]$ is used as a shorthand notation for the full expression in (\ref{rearrange})-(\ref{rearrange_EOM}).

Taking into account that $\hat O_1$ is an arbitrary operator, we get  
\begin{equation}
\frac{d}{dt} \mbox{Tr}_{LR}[\ldots ] = \mathcal{L}_{SA}\,\mbox{Tr}_{LR}[\ldots ]
\end{equation}
which has solution
\begin{equation}
\label{solved}
 \mbox{Tr}_{LR}[\ldots](t)=\mathcal{U}_{SA}(t,0)\, \mbox{Tr}_{LR}[\ldots](0) \equiv \mathcal{U}_{SA}(t,0)\, \hat O_2\,\hat\rho_{SA}(t_2)
\end{equation}
Substituting (\ref{solved}) into (\ref{rearrange}) leads to
\begin{equation}
 \label{corr2rt_solved}
 \bigg\langle\hat O_1(t+t_2)\,\hat O_2(t_2)\bigg\rangle = 
 \mbox{Tr}_{LR}\bigg[ \hat O_1\,\mathcal{U}_{SA}(t_1,t_2)\bigg(\hat O_2\,\hat\rho_{SA}(t_2)\bigg)\bigg]
\end{equation}
This relation expresses two-time correlation function defined from unitary evolution of the $aux$ model 
in terms of Lindblad-type evolution of $S+A$ subspace of the $aux$ model.
Finally, we note that while we had restriction $t>0$ in derivation of (\ref{EOM_green}), the result
is correct also for $t=0$, as one can see by direct comparison of the two sides in (\ref{corr2rt_solved}).
Eq.~(\ref{corr2rt_solved}) together with (\ref{rho_EOM}) leads to (\ref{corr2rt_EOM}).

Similarly, for $t_2\geq t_1\geq 0$ one can prove that
\begin{align}
 \label{single_green2}
& \langle\hat O_1(t_1)\hat O_2(t_2)\rangle = 
\\ &\qquad
\langle\langle I\vert \mathcal{O}^{-}_2\,\mathcal{U}_{SA}(t_1,t_2)\,
\mathcal{O}^{+}_{1}\,\mathcal{U}_{SA}(t_2,0)\,\vert\rho_{SA}(0)\rangle\rangle
\nonumber
\end{align}


\section{Derivation of Eq.~(\ref{corrN_EOM})}\label{appC}
Here we prove that multi-time correlation functions of arbitrary operators $\hat O_i$ in $S+A$ of the $aux$ model
defined on the Keldysh contour, 
\begin{equation}
\label{corrNdef}
 \bigg\langle T_c \hat O_1(\tau_1)\,\hat O_2(\tau_2)\,\ldots\,\hat O_N(\tau_N)\bigg\rangle,
\end{equation}
can be evaluated from Markov Lindblad-type evolution in the $S+A$ subspace.
Here operators $\hat O_i$ are in the Heisenberg picture. 
Projections (one-the-contour time orderings) of multi-time correlation functions (\ref{corrNdef}) 
will have the following form
\begin{align}
		& \bigg\langle\hat{B}_1(s_1)\hat{B}_2(s_2)..\hat{B}_m(s_m)\hat{C}_n(t_n)...\hat{C}_2(t_2)\hat{C}_1(t_1)\bigg\rangle
		\nonumber \\ &
		=\mbox{Tr}\bigg[\hat{C}_n(t_n)...\hat{C}_2(t_2)\hat{C}_1(t_1)\hat{\rho}^{aux}(0)
		\\ &\qquad\qquad\times
		                           \hat{B}_1(s_1)\hat{B}_2(s_2)...\hat{B}_m(s_m)\bigg]
		                           \nonumber
\end{align}
where $\hat{B}_j(s_j)$ and $\hat{C}_i(t_i)$ are used for operators $\hat O_i$ on the backward and forward branches of the contour, respectively (see Fig.~\ref{fig2}) and where
\begin{equation}
\begin{array}{l}
	t_n>t_{n-1}>...>t_1\geq 0
	\\
	s_m>s_{m-1}>...>s_1\geq 0
	\end{array}
\end{equation}
Note, there is no ordering between the sets $\{t_i\}$ and $\{s_j\}$ ($i\in\{1,2,\ldots,n\}$ and $j\in\{1,2,\ldots,m\}$).

Let denote the time-ordering of the set $\{t_1,t_2,...,t_n,s_1,s_2,...,s_m\}$ by 
$\{\theta_1,\ldots,\theta_{m+n}\}$. So that
\begin{equation}
	\theta_{m+n}\geq \theta_{m+n-1}\geq\ldots\geq \theta_1\geq0.
\end{equation}
We want to prove that projections of multi-time correlation functions satisfy quantum regression theorem~\cite{breuer_theory_2003}
\begin{align}
\label{QRT}
		&\langle\langle I \vert \mathcal{O}_{\theta_{m+n}}\,\mathcal{U}_{SA}(\theta_{m+n},\theta_{m+n-1})\,
		\mathcal{O}_{\theta_{m+n-1}}\,
		\\ &\qquad
		\mathcal{U}_{SA}(\theta_{m+n-1},\theta_{m+n-2})\ldots 	
		\mathcal{O}_{\theta_1}\, 
		\mathcal{U}_{SA}(\theta_1,0)\vert \rho_{SA}(0)\rangle\rangle
		\nonumber
\end{align}
where $\mathcal{O}_{\theta_i}$ is superoperator, Eq.~(\ref{superop}), corresponding to  operator $\hat B$  or $\hat C$
(backward or forward branch of the contour, respectively) at real time $\theta_i$. 

We prove (\ref{QRT}) by mathematical induction.
First, we note that Eqs.~(\ref{corr2rt_EOM}) and (\ref{single_green2}) are special cases of Eq.~(\ref{QRT}) with $m+n=2$. 
Suppose that for any combination $(m,n)$ satisfying $m+n=k$, Eq.~(\ref{QRT}) holds. 
Now let consider $(k+1)$-time correlation function
\begin{align}
	&\bigg\langle\hat{B}_1(s_1)\hat{B}_2(s_2)\ldots\hat{B}_m(s_m)\hat{O}_{\theta_{k+1}}(\theta_{k+1})\hat{C}_n(t_n)\ldots
	\\ &\qquad\qquad\qquad\qquad\qquad\qquad\qquad\ldots
	\hat{C}_2(t_2)\hat{C}_1(t_1)\bigg\rangle
	\nonumber
\end{align}
where $\theta_{k+1}>t_n>t_{n-1}>...>t_1\geq 0$ and $\theta_{k+1}>s_m>s_{m-1}>...>s_1\geq 0$. 
As previously, we time-order both sets,
\begin{equation}
	\theta_{k+1}>\theta_{k}\geq \theta_{k-1}\geq\ldots\geq \theta_1\geq 0,
	\end{equation}
and take the derivative with respect to the latest time
\begin{widetext}
\begin{align}
 \label{EOM_multi}
		&\frac{d}{d\theta_{k+1}}
		\bigg\langle\hat B_1(s_1)\,\hat B_2(s_2)\ldots\hat B_m(s_m)\,\hat{O}_{k+1}(\theta_{k+1})\,
		\hat C_n(t_n)\ldots\hat C_2(t_2)\,\hat C_1(t_1)\bigg\rangle
		\nonumber \\ 
		&\equiv \frac{d}{d\theta_{k+1}} \langle\langle I\vert \mathcal{O}_{\theta_{k+1}} \, \mathcal{U}^{aux}(\theta_{k+1},\theta_k)\,
		\mathcal{O}_{\theta_k}\,\mathcal{U}^{aux}(\theta_k,\theta_{k-1}) \ldots \mathcal{U}^{aux}(\theta_1,0)
		\vert \rho^{aux}(0)\rangle\rangle
		\\
		&= \mbox{Tr}_{SA}\bigg\{\hat O_{\theta_{k+1}} \frac{d}{d\theta_{k+1}}
		\langle\langle I_{LR}\vert \mathcal{U}^{aux}(\theta_{k+1},\theta_k)\,\mathcal{O}_{\theta_k}\,\mathcal{U}^{aux}(\theta_k,\theta_{k-1})
		\ldots \mathcal{U}^{aux}(\theta_1,0)\vert \rho^{aux}(0)\rangle\rangle_{LR}\bigg\}
		\nonumber \\
		&= \mbox{Tr}_{SA}\bigg\{\hat O_{\theta_{k+1}}
		\mathcal{L}_{SA}(\theta_{k+1})\,
		\langle\langle I_{LR}\vert \mathcal{U}^{aux}(\theta_{k+1},\theta_k)\,\mathcal{O}_{\theta_k}\,\mathcal{U}^{aux}(\theta_k,\theta_{k-1})
		\ldots \mathcal{U}^{aux}(\theta_1,0)\vert \rho^{aux}(0)\rangle\rangle_{LR}\bigg\}
		\nonumber
\end{align}
where we followed the argument leading to (\ref{rearrange_EOM}) and (\ref{EOM_green_rewritten}) in \ref{appB}.
In (\ref{EOM_multi}) $\mathcal{U}^{aux}$ is the Liouville space analog of the Hilbert space evolution operator $\hat U^{aux}$
defined in Eq.~(\ref{Uaux}).

Taking into account that $\hat O_{\theta_1}$ is an arbitrary operator, we get
\begin{align}
\label{2solve}
 &\frac{d}{d\theta_{k+1}} \langle\langle I_{LR}\vert \mathcal{U}^{aux}(\theta_{k+1},0)\ldots \vert \rho^{aux}(0)\rangle\rangle_{LR}
  = \mathcal{L}_{SA}(\theta_{k+1})\, \langle\langle I_{LR}\vert \mathcal{U}^{aux}(\theta_{k+1},0)\ldots \vert \rho^{aux}(0)\rangle\rangle_{LR}
\end{align}
where $\langle\langle I_{LR}\vert \mathcal{U}^{aux}(\theta_{k+1},0)\ldots \vert \rho^{aux}(0)\rangle\rangle_{LR}$ 
is shorthand notation for the expression introduced in (\ref{EOM_multi}). 

Solving (\ref{2solve}) and utilizing quantum regression theorem for its initial condition, $\theta_{k+1}=\theta_k$,
leads to 
\begin{align}
\label{induction}
 &\bigg\langle\hat B_1(s_1)\,\hat B_2(s_2)\ldots\hat B_m(s_m)\,\hat{O}_{\theta_{k+1}}(\theta_{k+1})\,
		\hat C_n(t_n)\ldots\hat C_2(t_2)\,\hat C_1(t_1)\bigg\rangle
\\ &\quad
	=\langle\langle I\vert \mathcal{O}_{\theta_{k+1}}\,\mathcal{U}_{SA}(t_{\theta_{k+1}},t_{\theta_{k}})\, \mathcal{O}_{\theta_{k}}\,
 \mathcal{U}_{SA}(t_{\theta_{k}},t_{\theta_{k-1}})\ldots \mathcal{O}_{\theta_1}\mathcal{U}_{SA}(t_{\theta_1},0)
 \vert \rho_{SA}(0)\rangle\rangle
 \nonumber
\end{align} 
which is quantum regression theorem for $(k+1)$-time correlation function.
\end{widetext}
Thus,  by induction we prove Eq.~(\ref{corrN_EOM}).


\end{document}